\input harvmac.tex
\let\includefigures=\iffalse
\let\useblackboard=\iffalse
\def\Title#1#2{\rightline{#1}
\ifx\answ\bigans\nopagenumbers\pageno0\vskip1in%
\baselineskip 15pt plus 1pt minus 1pt
\else
\def\listrefs{\footatend\vskip 1in\immediate\closeout\rfile\writestoppt
\baselineskip=14pt\centerline{{\bf References}}\bigskip{\frenchspacing%
\parindent=20pt\escapechar=` \input
refs.tmp\vfill\eject}\nonfrenchspacing}
\pageno1\vskip.8in\fi \centerline{\titlefont #2}\vskip .5in}

\ifx\answ\bigans\def\tcbreak#1{}\else\def\tcbreak#1{\cr&{#1}}\fi
%

%

\def\comments#1{}

\def\p{\partial}

\def\half{{1\over 2}}

\def\vev#1{\langle{#1}\rangle}

\def\CM{{\cal M}}
\def\CN{{\cal N}}

\def\nl{\hfill\break}

\def\I{I}

\def\II{\relax{I\kern-.10em I}}
\def\IIa{{\II}a}

\def\IZ{\relax\ifmmode\mathchoice
{\hbox{\cmss Z\kern-.4em Z}}{\hbox{\cmss Z\kern-.4em Z}}
{\lower.9pt\hbox{\cmsss Z\kern-.4em Z}}
{\lower1.2pt\hbox{\cmsss Z\kern-.4em Z}}\else{\cmss Z\kern-.4em
Z}\fi}
\def\IB{\relax{\rm I\kern-.18em B}}
\def\IC{{\relax\hbox{$\inbar\kern-.3em{\rm C}$}}}
\def\ID{\relax{\rm I\kern-.18em D}}
\def\IE{\relax{\rm I\kern-.18em E}}
\def\IF{\relax{\rm I\kern-.18em F}}
\def\IG{\relax\hbox{$\inbar\kern-.3em{\rm G}$}}
\def\IGa{\relax\hbox{${\rm I}\kern-.18em\Gamma$}}
\def\IH{\relax{\rm I\kern-.18em H}}
\def\II{\relax{\rm I\kern-.18em I}}
\def\IK{\relax{\rm I\kern-.18em K}}
\def\IP{\relax{\rm I\kern-.18em P}}

%
\def\mod{{\rm mod}}

\def\p{\partial}

\font\cmss=cmss10 \font\cmsss=cmss10 at 7pt
\def\IR{\relax{\rm I\kern-.18em R}}

\def\inbar{\,\vrule height1.5ex width.4pt depth0pt}

\def\BR{\IR}
\def\BZ{\IZ}
\def\BR{\IR}
\def\BC{\IC}

\def\lp10{l_P^{10}}
\def\lp11{l_P^{11}}
\def\R11{R_{11}}

\Title{\vbox{\baselineskip14pt\hbox{hep-th/9612126}
\hbox{RU-96-111}}}
{\vbox{\baselineskip20pt
\centerline{Enhanced Gauge Symmetry}
\centerline{in M(atrix) Theory} }}
\centerline{Michael R. Douglas}
\medskip
\centerline{\it Department of Physics and Astronomy}
\centerline{\it Rutgers University }
\centerline{\it Piscataway, NJ 08855--0849}
\medskip
\centerline{\tt mrd@physics.rutgers.edu}
\medskip
\bigskip
\noindent
We discuss the origin of enhanced gauge symmetry in ALE (and K3)
compactification
of M theory, either defined as the strong coupling limit of the type
\IIa\ superstring, or as defined by Banks et al.
In the D-brane formalism, wrapped membranes are D0 branes with twisted
string boundary conditions, and appear on the same footing with
the Kaluza-Klein excitations of the gauge bosons.
In M(atrix) theory, the construction appears to work for arbitrary
ALE metric.
\Date{December 1996}
\def\half{{1 \over 2}}

\def\laplace{{\kern1pt\vbox{\hrule height 1.2pt\hbox{\vrule width 1.2pt\hskip
  3pt\vbox{\vskip 6pt}\hskip 3pt\vrule width 0.6pt}\hrule height 0.6pt}
  \kern1pt}}
\def\scriptlap{{\kern1pt\vbox{\hrule height 0.8pt\hbox{\vrule width 0.8pt
  \hskip2pt\vbox{\vskip 4pt}\hskip 2pt\vrule width 0.4pt}\hrule height 0.4pt}
  \kern1pt}}

\lref\BFSS{T. Banks, W. Fischler, S. H. Shenker and L. Susskind,
``M Theory As A Matrix Model: A Conjecture,'' hep-th/9610043.}
\lref\bd{M. Berkooz and M. R. Douglas, ``Five-branes in M(atrix) Theory,''
hep-th/9610236.}
\lref\suss{L. Susskind,
``T Duality in M(atrix) Theory and S Duality in Field Theory,''
hep-th/9611164.}
\lref\grt{O. Ganor, S. Rangoolam and W. Taylor,
``Branes, Fluxes and Duality in M(atrix)-Theory,'' hep-th/9611202.}
\lref\aha{O. Aharony and M. Berkooz,
``Membrane Dynamics in M(atrix) Theory,'' hep-th/9611215.}
\lref\hulltown{C. Hull and P. Townsend, Nucl. Phys. B438 (1995) 109;
hep-th/9410167.}
\lref\witten{E. Witten, Nucl. Phys. B443 (1995) 85, hep-th/9503124.}
\lref\McKay{J. McKay, Proc. Symp. Pure Math vol. 37, AMS (1980) 183.}
\lref\kron{P. Kronheimer, ``The construction of ALE spaces as
hyper-kahler quotients,'' J. Diff. Geom. {\bf 29} (1989) 665.}
\lref\kn{P.B. Kronheimer and H. Nakajima, ``Yang-Mills instantons
on ALE gravitational instantons,'' Math. Ann. {\bf 288} (1990) 263.}
\lref\dm{M. R. Douglas and G. Moore, ``D-Branes, Quivers, and ALE Instantons,''
hep-th/9603167.}
\lref\jm{C. Johnson and R. Myers, ``Aspects of Type IIB Theory on ALE Spaces,''
hep-th/9610140.}
\lref\aspinwall{P. S. Aspinwall, Phys. Lett. {\bf B357} (1995) 329-334;
hep-th/9507012.}
\lref\joerev{
S.~Chaudhuri, C.~Johnson, and J.~Polchinski,
hep-th/9602052; J.~Polchinski, ``TASI Lectures on D-Branes,'' hep-th/9611050.}
\lref\polpro{J.~Polchinski, ``Tensors from K3 Orientifolds,''
hep-th/9606165.}
\lref\dmunp{M. R. Douglas and G. Moore, unpublished.}
\lref\toap{M. R. Douglas, S. H. Shenker and L. Susskind, work in progress.}
\lref\hm{J. A. Harvey and G. Moore, ``On the algebra of BPS states,''
hep-th/9609017.}
\lref\naka{H. Nakajima, ``Homology of moduli
spaces of instantons on ALE Spaces. I'' J. Diff. Geom.
{\bf 40}(1990) 105; ``Instantons on ALE spaces,
quiver varieties, and Kac-Moody algebras,'' Duke Math.
{\bf 76} (1994)365;
``Gauge theory on resolutions of simple singularities
and simple Lie algebras,'' Intl. Math. Res. Not.
{\bf 2}(1994) 61;
``Quiver Varieties and Kac-Moody algebras,''
preprint; ``Heisenberg algebra and Hilbert
schemes of points on projective surfaces,''
alg-geom/9507012; ``Instantons and
affine Lie algebras,'' alg-geom/9502013.}
\lref\dewit{B. de Wit, J. Hoppe and H. Nicolai, Nucl.~Phys. B305(1988)
 545.\nl
B. de Wit, M. Luscher and H. Nicolai, Nucl.~Phys. B320(1989) 135.}
\lref\town{P. Townsend, Phys. Lett. B373 (1996) 68;
hep-th/9512062.}
\lref\dzero{U.H. Danielsson, G. Ferretti and  B. Sundborg,
``D-particle Dynamics and Bound States'', hep-th/9603081\nl
D. Kabat and P. Pouliot, ``A Comment on Zero-Brane Quantum
Mechanics'', hep-th/9603127}
\lref\DKPS{M. R. Douglas, D. Kabat, P. Pouliot and S. Shenker,
``D-branes and Short Distances in String Theory,'' hep-th/9608024.}
\lref\shenker{S.~H.~Shenker, ``Another Length Scale in String Theory?,''
hep-th/9509132.}
\lref\PolRR{J.~Polchinski, Phys.~Rev.~Lett.~{\bf 75} (1995) 4724,
hep-th/9510017.}
\lref\kp{D.~Kabat and P.~Pouliot,
{\it A Comment on Zero-Brane Quantum Mechanics}, hep-th/9603127.}
\lref\guven{R. Gueven, Phys. Lett. B276 (1992) 49.}
\lref\duff{M. J. Duff, ``M Theory,'' hep-th/9608117.}
\lref\sethi{S.~Sethi and M.~Stern, ``A Comment
on the Spectrum of H-Monopoles,'' hep-th/9607145.}
\lref\sen{A. Sen, Phys.Rev. D54 (1996) 2964, hep-th/9510229.}

\newsec{Introduction}

Recently Banks et. al. have proposed a definition of
eleven-dimensional M theory in the infinite momentum frame,
as a large $N$ limit of a supersymmetric
matrix quantum mechanics \BFSS.
To support this, they start with the fact that this quantum mechanics
describes the D$0$-branes which dominate the strong coupling
limit of \IIa\ string theory, argue that anti D$0$-branes decouple in the
IMF,
and then adapt results of \DKPS\
showing that the theory can reproduce supergravity interactions
without need of the original closed strings.
{}From this point of view, modifications to the background can be made by
adapting the corresponding modifications to the \IIa\ string, providing
definitions of the five-brane \bd\ and toroidal compactifications
\refs{\BFSS,\suss,\grt}.
Not all physics follows from \IIa\ arguments however and
a quite non-trivial non-\IIa\ result is the appearance of the supermembrane
with correct physics
\refs{\dewit,\town,\BFSS,\aha}.

In this note we study another example of \IIa--derived M-theory physics;
the enhanced gauge symmetry of
compactifications on $K3\times \BR^7$ in the orbifold limit \witten\ which
follows from the proposal by Hull and Townsend of strong-weak
coupling duality between \IIa\ on $K3$ and the heterotic
string on $T^4$ \hulltown.
Its M theory origin is clear --
membranes wrapped on small supersymmetric two-cycles become particles
with conventional (vector) gauge charge in the dimensionally reduced theory,
and when such two-cycles degenerate to zero volume, these particles
include massless gauge bosons.

Since this phenomenon is local, we can see it by formulating the theory in
the neighborhood of the degenerating two-cycles, in other words on
$\CM_\zeta\times\BR^7$, where $\CM_\zeta$ is an ALE space
asymptotic to $\BC^2/\Gamma$.
$\Gamma$ is a finite subgroup of $SU(2)$
and it has an associated simply laced extended
Dynkin diagram $\CG$, an affine Lie algebra $\hat G$ and finite Lie algebra
$G$ \McKay.
The enhanced gauge symmetry obtained by maximal degeneration
to the singularity $\BC^2/\Gamma$ is simply $G$.

An explicit hyperk\"ahler quotient construction
of $\CM_\zeta$ with its metric was made by Kronheimer \kron,
and this construction appears in D-brane physics:
the natural construction of D-branes embedded at a point in an orbifold
produces a gauge theory whose moduli space is $\CM_\zeta$ \dm.
We will use this construction for D$0$-branes in \IIa\ string theory.

A wrapped D$2$-brane also has a known D-brane realization
in this construction \refs{\polpro,\dmunp}:
it is a D$0$-brane with twisted boundary conditions for the
open strings, which project out the moduli moving it from the fixed point.
An easy computation shows that it is charged under
twist sector RR fields, and since the wrapped membrane is the only
charged BPS state in the large volume limit, the two objects must
be continuously connected.
Its non-zero mass is
interpreted as a consequence of an implicit
$B \ne 0$ of the orbifold construction \aspinwall;
we will be able to determine $B$ for any $\Gamma$.

This $B$ corresponds to a Wilson line in the additional dimension of
M theory, and thus the Kaluza-Klein states of massless gauge bosons and the
massive gauge bosons of spontaneously broken gauge symmetry appear on the
same footing in this construction.

Taking this construction for D$0$-branes and adapting it
according to the rules of
\BFSS\ provides a construction of M theory on $\CM_\zeta\times\BR^7$.
There are several differences with the string theory discussion.
First, the moduli of the ALE are controlled by Fayet-Iliopoulos terms
in the gauge theory.  While these were derived
in \dm\ as couplings to closed string twist fields,
here they are postulated.  This is appropriate as we are
discussing different backgrounds in the infinite momentum frame, which
should be realized by changing parameters in the Lagrangian.
Second, the expectation value of $B$ disappears in
the limit, and the full enhanced gauge symmetry appears.
Finally, there appears to be no analog of the upper bound on the blow-up
parameter $\zeta$ at the string scale which follows from the general
results of \DKPS.

Another test can be made by introducing a five-brane
wrapped on K3 or $M_\zeta$.  This produces the heterotic string
which dominates the strong coupling limit, and we must see a level $1$ action
of $\hat G$ on the spectrum of this string.

As in \bd, we define the five-brane by introducing a vector hypermultiplet
into the D$0$-brane quantum mechanics.  This theory
is the dimensional reduction of the general theory of \dm,
corresponding to the hyperk\"ahler quotient construction of
instanton moduli space of Kronheimer and Nakajima \kn.
Following Harvey and Moore
\hm, if we assume that the space of bound states is the
sheaf cohomology of this moduli space,
then results of Nakajima \naka\
imply the existence of these bound states as well as the $\hat G$ action.

\newsec{D$0$-branes on orbifolds}

D-branes on an orbifold are defined as in \dm; we take
the $U(N)$ gauge theory of $N$ D-branes at the fixed point in $\BC^2$
and quotient by a combined action of $\Gamma$
on space-time and the Chan-Paton factors:
$A_\mu = \gamma_{CP} A_\mu \gamma_{CP}^{-1}$
and $Z^i = \gamma^i_j \gamma_{CP} Z^j \gamma_{CP}^{-1}$.
The derivation can be made for $5$-branes and the result is
a $\CN=1$, $d=6$ gauge theory whose Lagrangian (at leading
order in $\alpha'$, which is the only part used in \BFSS)
is determined by the choice of gauge group and matter representation;
this is given in \dm\ for the A series, and
in \jm\ for the D and E series.
The quantum mechanics of D$0$-branes is its dimensional reduction.

The field content is determined by a choice of $\Gamma$ representation $R$.
Let the irreducible representations of $\Gamma$ be $R_i$
with $0\le i\le {\rm rank}\ G$ and their dimensions be $n_i$.
$R_0$ is the trivial representation, $R_1$ the fundamental (the
same as the action on $\BC^2$), and the $R_i$ are
associated with the extended Cartan matrix $\hat C$ by the McKay
correspondance,
\eqn\mckay{
R_1 \otimes R_i = \oplus_j (2\delta -\hat C)_{ij} R_j.
}
In terms of the extended Dynkin diagram $\CG$, each node is an irrep $R_i$,
and the non-zero terms in \mckay\ are links.
If $R$ decomposes as
\eqn\Rsum{R = \sum_{i=0}^r v_i R_i,}
the resulting gauge symmetry is $\prod_i U(v_i)$, and each link of $\CG$
comes with a hypermultiplet in the $(v_i,\bar v_j)$.
Besides the overall coupling, the parameters of the Lagrangian are
an $SU(2)_R$ triplet of Fayet-Iliopoulos terms
for each of the $U(1)$ factors, $\zeta_i^n$ with $1\le n\le 3$,
$0\le i\le r$ and $\sum_i \zeta_i^n = 0$.
These determine the periods of the metric on the hyperk\"ahler quotient
and in string theory are determined by expectations
of twist fields \dm.

The simplest case to interpret is $N$ copies of the regular
representation, $v_i=N n_i$.  The Higgs branch of the moduli
space for generic $\zeta$ is the symmetric product
$(\CM_\zeta\times \BR^5)^N/S_N$,
the positions of $N$ independent D$0$-branes in space-time.
Following \witten,
these are interpreted as Kaluza-Klein states of M theory
on $\CM_\zeta\times \BR^7$ with momentum $p_{11}=1/R_{11}$.
Consistency of this interpretation predicts bound states with
$p_{11}=N/R_{11}$ corresponding to all partitions of this momenta among
up to $N$ particles.

Along with the bulk supergravity fields,
we expect additional bound states localized near the fixed point;
in particular the BPS states which come from the decomposition
\eqn\decomp{
C^{(3)} = \sum_i A^{(i)}_\mu(x) \omega^{(i)},
}
where $\omega^{(i)}$ are the normalizable harmonic forms on the ALE,
and their supersymmetry partners forming a full gauge multiplet.

The simplest case in which to check this is $\Gamma=\BZ_2$, $v_1=v_2=1$.
This is $U(1)^2$ gauge theory, but the diagonal $U(1)$ decouples, and
the non-trivial dynamics is that of $U(1)$ gauge theory coupled to two
hypermultiplets of charge $2$.  This is a system for which the arguments of
\refs{\sen,\sethi,\DKPS} establish the existence of bound states;
it is identical
(up to the unit of charge) to the system of a D$0$-brane in the presence of
two D$4$-branes and using string duality, has the same bound states as two
D$0$-branes and a single D$4$-brane.  Thus one predicts
a ``new'' bound state and an additional set of BPS states in the symmetric
product of two single 0--4 bound states.  The new bound state will be
interpreted as the $p_{11}=1/R$ KK mode of the $U(1)$ gauge multiplet,
while the others could be interpreted as the product of a
bound state with $(v_1,v_2)=(1,0)$ and one with $(v_1,v_2)=(0,1)$.

Now there was no consistency condition in the string theory requiring
all $v_i$ equal, and the interpretation of more general states is
briefly described in \polpro\ (it was known to the authors
of \dm\ but not mentioned there):
taking a single $v_i=1$ and the rest zero produces a D$2$-brane wrapped
around a non-trivial two-cycle of the ALE.  In string theory,
the test of this is to check that it is a source of twist sector RR field,
the orbifold realization of the fields \decomp.  This is shown in the
appendix to \dm, for $\Gamma=\BZ_n$.

The representations $R_i$ are associated with homology two-cycles $\sigma_i$
in $\CM_\zeta$, and $R_0$ is associated with $\sigma_0=-\sum_i n_i\sigma_i$.
The intersection form $\vev{\sigma_i \cup \sigma_j}$ is the
extended Cartan matrix $\hat C_{ij}$ \kron.
This leads to a further association of $R_i$ and $\sigma_i$ with
the simple roots $\vec \alpha_i$ and the lowest root
$\vec \alpha_0$ of $G$; these translate directly into the charges
$\vec Q_i=\vec \alpha_i$
of the $r+1$ elementary wrapped two-branes.

The coupling to the untwisted sector is universal, so all of these D$0$-branes
have mass $1/c_2 g_s$, where $c_2=\sum_i n_i$ is the Coxeter number
of $\Gamma$.
This mass is also determined by the central charge
formula \hulltown\ to be
\eqn\ccmass{
m={1\over g_s}\min_{n\in \BZ} |\int_{\vec Q\cdot\vec\sigma} B+iJ\ +n|
}
where $J$ is the K\"ahler form with respect to the complex structure
for which $\sigma$ is a holomorphic curve (i.e. with $\int \Omega=0$).
Matching this for $\zeta=0$
determines the background $B$ for the orbifold.
\footnote*{The dependence of the mass of a finite
wrapped D2-brane on $B \mod 1$
is realized by the additional
dependence on a world-volume gauge field $F$ and
the possibility of $\int F\ne 0$.  In the present definition, it
appears to be reflected in subtleties in the D$0$-brane coupling to the
twist sector $B$ similar to those found in the appendix of \dm.}
For $\BZ_2$, it is $B=1/2$ as found in \aspinwall.
For $\BZ_n$, the cycles $\sigma_i, i>0$ associated
with simple roots have $\int B = 1/n$.  This corresponds to a non-zero
$C_{11\mu\nu}$ in eleven dimensions and thus to an $SU(n)$ Wilson line
\eqn\willine{
A_{11} = {1\over R_{11}}\left(\matrix{
{\sigma\over n}&0&0&\ldots&0\cr
0&{\sigma+1\over n}&0&\ldots&0\cr
0&0&{\sigma+2\over n}&\ldots&0\cr
\vdots&\vdots&\vdots&\ddots&\vdots\cr
0&0&0&\ldots&{\sigma-1\over n}}\right)
}
where $\sigma={n-1 \over 2}$.
There is a strong analogy with the Wilson line breaking $E_8$ to $SO(16)$
in the relation of type \I' string theory to M theory \joerev.
Perhaps there is a general rule determining such symmetry breakings.

Thus the $\sum v_i=1$ states provide the gauge bosons corresponding to
simple roots of $G$.  They fall into $8+8$ component supermultiplets;
more explicitly the orbifold projection retains half of the $16$
components of the gaugino, transforming in the doublet of the
$SU(2)\subset SO(4)$ which is singlet under the orbifold projection
and the $4$ of $SO(5)$.  These act on a multiplet whose
bosons are a vector of $SO(5)$ and three scalars, the physical states of
a $d=7$ gauge boson and the metric fluctuations.

Their bound states must provide all of the gauge bosons, and
thus we predict that a new supermultiplet of bound
states exists for each root $\alpha$ and integer $N$, with
$p_{11}=(N+\half B\cdot \alpha)/R_{11}$; in other words for
each sector with $(\sum_i v_i \alpha_i)^2 = 2$.
This must be true in the full \IIa\ theory for consistency
of M theory; the
explicit bound state we described lends support to the conjecture
that these are bound states in the pure D$0$ quantum mechanics.

Although enhanced gauge symmetry is spontaneously broken,
it is fairly manifest.
In the D-brane realization, supergravity Kaluza-Klein states and
wrapped membrane states appear on an equal footing.

Turning on the moduli $\zeta_i^n$ modifies the gauge symmetry breaking
(they correspond to Wilson lines $A_n$ in the
$T^3$ of the dual heterotic string)
and the effective Hamiltonian.  For $|\zeta|<<\vev{B}$, \ccmass\ has the
expansion
\eqn\ccmassexp{\eqalign{
m &= {1\over g_s}\sqrt{B^2 + \zeta^2} \cr
&= {B\over g_s} + {\zeta^2\over 2B g_s} + O(\zeta^4)
}}
where $\zeta^2=\sum_n(\vec Q\cdot\vec\zeta^n)^2$.
The $O(\zeta^2)$ dependence can be seen explicitly for $\sum v_i=1$
in the D-term potential,
which degenerates to $V \propto \sum_n(\zeta^n)^2$.  For
$\zeta\sim 1$, stringy corrections are known to be important \dm.

\subsec{From M theory to M(atrix) theory}

Following \BFSS, we now regard this system as the definition of M theory
on $\CM_\zeta$ in a sector with longitudinal light-cone
momentum $P_{-}=N/R$, and take the $R_{11}\rightarrow\infty$ limit.
The Wilson line \willine\ disappears, and the massless charged
gauge bosons at $\zeta=0$ are manifest.  Now it is this observation which
confirms the identification of $\sum v_i=1$ states as wrapped membranes.

It is important to check the basic tenets of \BFSS\ in this context,
for example that supergravity interactions
between these particles are correctly reproduced by quantum open string
effects.
This issue will be discussed in \toap.

For any $|\zeta| >> l_{p11}^2$, the classical analysis of the
resulting Higgs branch appears to be valid, meaning there would be
no restriction on the blow-up parameter in this construction.
Consistent with this, the D-term potential exactly reproduces the term
$m^2/p_{11}=\zeta^2/p_{11}$
in the IMF $0$-brane energy.

The construction must work for all states, not just BPS states.
There are clear predictions for the states on the blowup with $\zeta>>l_p^2$,
where the conventional supergravity analysis is valid: we
diagonalize the basic supergravity and membrane Hamiltonians
to get higher modes in the KK expansion \decomp,
and local excitations of the wrapped membranes.
However, estimating their couplings to the bulk states
using the known membrane coupling $\int h_{\mu\nu} \p X^\mu \p X^\nu$
leads to the conclusion that they are unstable and thus only
the full dynamics can be sensibly compared, a very interesting open problem.

\newsec{Five-branes}

We add a five-brane as in \bd, by adding vector degrees of freedom.
Different orientations will have different physics.
We can put it at a point in the ALE (by starting off with images),
and get a theory which should contain ``tensionless strings'' in the
limit $\zeta=X=0$. Seeing these should be quite interesting but
requires knowing how to construct membranes ending on the
five-branes.

If we instead
embed the longitudinal dimensions in an ALE, we get a piece of
the five-brane wrapped around K3.
This is the heterotic soliton which dominates the
small K3 limit.  We are treating the large K3 limit, but we must see BPS
states of this soliton in any case.  These are excitations of the bosonic
left movers admitting unbroken $(0,4)$ supersymmetry and
the action of world-sheet current
algebra, affine $\hat G$ at level $1$.  This symmetry
is broken both by
$\zeta\ne 0$ and, at finite $R_{11}$, by the Wilson line \willine, but
this will be realized by explicit terms in the Hamiltonian.

Physical five-brane degrees of freedom are new bound states
of zero-branes.  In the IMF, we identify the left and right world-sheet stress
tensors with
\eqn\hetident{\eqalign{
H &= L_0 = \half (p-w)^2 + N_0 \cr
P_{11} &= \bar L_0 = \half (p+w)^2 + \bar N_0.
}}
Note that we are not restricted to $L_0=\bar L_0$, because we are considering
a finite piece of an infinite string.

The choice of which chirality has world-sheet supersymmetry is
determined by the chirality of the additional vector degrees of freedom.
If we make this compatible with the unbroken supersymmetry on the orbifold,
we get non-trivial supersymmetries commuting to produce $H$, so $L_0$
is the supersymmetric side (say right movers) and BPS excitations can have
non-zero $\bar N_0$.  If we make the other choice, the supersymmetry becomes
trivial and $\bar L_0$ is the supersymmetric side.  In this case, BPS
states will not be realized as bound states of D$0$-branes.


The full gauge theory is now a D$0$--D$4$ brane system, also derived
in \dm\ (section 5).  These theories are parameterized by a set of
non-negative integers $w_i$ where $\sum w_i$
is the total number of D$4$-branes.
They are obtained from the pure D$0$-brane theories
by adding $w_i$ hypermultiplets in the fundamental
representation of $U(v_i)$ for each $i$.
As shown in \kn\ (and reviewed in section 9 of \dm),
the Higgs branch of moduli space is generally
equivalent to a moduli space of instantons in the D$4$-brane gauge theory.
The choice of $w_i$ translates into a choice of first Chern class in this
language; a single heterotic string would have a single $w_k=1$, with $k$
denoting a choice of sector in the world-sheet theory.

It is natural to look for D$0$--D$4$ bound states in
the supersymmetric quantum mechanics on this moduli space, and thus
identify them with elements of the moduli space
cohomology.  Of course the moduli space approximation is not exact and
furthermore these spaces are typically singular.
Harvey and Moore \hm\ discuss some of the issues here, and propose that
the general identification will be between the Hilbert space of bound states
and the complex cohomology of the moduli space of coherent
simple sheaves.  This generalization is particularly significant in the
present case of a single D$4$-brane, as ``$U(1)$ self-dual instantons''
are at best rather singular objects.

Existing results on bound states in quantum mechanics along with the
string duality arguments of \hm\ all support this identification, and
we will assume it here.
This allows us to make use of the results of Nakajima \naka\ on the
cohomology and especially the celebrated Kac-Moody algebra which
acts on the cohomology.  For our present case of
$\sum_i w_i = 1$, this will be a $\hat G$ action
at level $1$.  The generators of this algebra $E_i$, $F_i$ and $H_i$ are
as follows:
the Cartan subalgebra $H_i$ acting on a cohomology class of
the sector of moduli space characterized by integers $v_i$ has eigenvalue
$v_i$; the operators $E_i$ add a single twisted D$0$-brane
(and thus increase $v_i$ by one); the
operators $F_i$ are their conjugates.  These are effectively
`second quantized' operators and their natural physical interpretation is
in terms of Harvey and Moore's ``correspondance conjecture'' \hm,
defining their action on the BPS states.

We claim that this is the standard world-sheet current algebra
which acts on the spectrum of a single heterotic string.
One test of this is that the left-moving Virasoro generators $\bar L_n$
must contain the Sugawara stress-tensor as one component.
This requires that $P_{11}$ as defined in M theory, i.e.
$(\sum v_i+c_k)/c_2 R_{11}$
where $c_k$ is a constant possibly depending on the sector $k$, be equal
to the Sugawara $L_0$.
This follows from Nakajima's results, which make $L_0$ the second Chern
class of the sheaf.  The constant $c_k$ is a contribution from the non-zero
first Chern class present for $w_i>0$, $i\ne 0$.

Nakajima's results also support this identification of the spectrum
-- in particular, it is shown
that the cohomology contains all highest weight representations --
but we have not verified
that the full cohomology is isomorphic to the spectrum of BPS states.
Following Harvey and Moore, this must follow from
\IIa\ -- heterotic string duality, because the D$0$ and D$4$ branes
of the construction are sensible objects in the \IIa\ string.
Indeed, the present discussion differs from theirs mainly in that we
are considering a state containing an infinitely long heterotic string
rather than perturbative heterotic string states.

\newsec{Conclusions}

In this note we showed that Dirichlet branes on orbifolds provide
a simple and explicit way to see the enhanced gauge symmetry of the \IIa\
string and M theory on K3.  The construction produces an explicit
realization of world-sheet current algebra for the wrapped five-brane which
becomes the dual heterotic string.  Although these are
not first quantized operators (they change the D$0$-brane number), it should
be possible to describe their action fairly explicitly.

In principle, the same operators adding and removing
zero-branes act on the space of pure $0$-brane bound states.
They will realize the subgroup of global gauge transformations and
it might be (extrapolating beyond Nakajima's results)
that this is contained in a Kac-Moody algebra
at level zero.  The natural interpretation of such an algebra
in  M theory (with $X^{11}$ space-like, so before going to the IMF)
would be the subgroup of gauge transformations
with $X^{11}$ dependence.
Such an interpretation would imply that the
Kac-Moody action can be extended to all states, not just BPS states.

The Virasoro algebra associated with Nakajima's Kac-Moody algebra is of
course part of the Virasoro algebra which plays the key role in
the Lorentz invariance of the light-cone heterotic string.
We believe that extending this action to the full state space will be a
key element in understanding the Lorentz invariance not manifest in the
treatment of \BFSS.

\medskip
We acknowledge valuable conversations with
P. Aspinwall, T. Banks, M. Berkooz, G. Moore, H. Nakajima and
S. Shenker.

This work was supported in part by DOE grant DE-FG02-96ER40559.
\listrefs
\end